\documentclass[aps,prd,nofootinbib,eqsecnum,notitlepage,10pt]{revtex4-1}
\usepackage{amssymb}
\usepackage{amsmath}
\usepackage{amsthm} 
\usepackage{mathrsfs}
\usepackage{amsfonts}
\usepackage{latexsym}
\usepackage{physics}
\usepackage{calc}
\usepackage{leftidx}
\usepackage{multirow}
\usepackage{url}
\usepackage{graphicx}
\usepackage{comment}
\usepackage{float}
\usepackage{caption}
\usepackage{subcaption}
\usepackage{enumerate}
\usepackage{csquotes}
\usepackage{graphics}
\usepackage{graphicx}
\usepackage{color}
\renewcommand{\thesection}{\arabic{section}}
\renewcommand{\thesubsection}{\thesection.\arabic{subsection}}
\renewcommand{\thesubsubsection}{\thesubsection.\arabic{subsubsection}}
\numberwithin{equation}{section}

\newcommand{\be}{\begin{equation}}
\newcommand{\ee}{\end{equation}}
\newcommand{\bes}{\begin{subequations}}
\newcommand{\ees}{\end{subequations}}
\def\bal#1\eal{\begin{align}#1\end{align}}

\usepackage[hidelinks]{hyperref}
\hypersetup{
    pdfnewwindow=true,
    plainpages=false,
}
\begin{document}
\title{Quantization of inhomogeneous spacetimes with cosmological constant term}
\author{Adamantia Zampeli}
\email{azampeli@phys.uoa.gr}
\affiliation{Department of History and Philosophy of Science, National and Kapodistrian University of Athens, 15771, Panepistimiopolis, Greece}
\affiliation{Institute of Theoretical Physics, Faculty of Mathematics and Physics, Charles University, V Hole\v{s}ovi\v{c}k\'ach 2, 18000 Prague 8, Czech Republic}

\author{Andronikos Paliathanasis}
\email{anpaliat@phys.uoa.gr}
\affiliation{Institute of Systems Science, Durban University of Technology, Durban 4000,
South Africa}

\begin{abstract}
We show that the Szekeres system with cosmological constant admits a sufficient number of conservation laws, which allow to claim the integrability of the system. The main novelty in this investigation is that we find that the unique attractor of the Szekeres system is the isotropic inhomogeneous de Sitter (-like) universe, contrary to the original system in which the attractors describe Kantowski-Sachs (-like) spacetimes. We also study the existence of quantum corrections and the emergence of classicality by considering the linear and quadratic conserved quantities at the quantum level. We perform an analysis considering different approaches, involving the Bohmian quantum potential and a probability analysis. The result is that there are no quantum corrections for the quadratic integrals, while there exists a linear case for which we find quantum corrections.
\end{abstract}

\maketitle
\section{Introduction}
Exact gravitational solutions have attracted much attention since the formulation of general relativity and several mathematical methods have been employed to advance the discovery of new solutions. Of particular interest are solutions which describe black holes and the universe at large scales. For the latter, the spatially homogeneous spacetimes is an important class, due to the present observed spatial homogeneity at the large scales. Still, though, there are observational deviations from strict homogeneity, a fact which opens many possibilities for the structure of the early universe during and after the Planck era. Indeed, inhomogeneous cosmological models constitute an alternative explanation of the small inhomogeneities in the cosmic microwave background \cite{AndrzejKrasinski:1997zz,Bolejko:2008xh,Ishak:2011hz}, but they can also provide alternative geometric mechanics in order to explain the inflationary era \cite{Clough:2016ymm,East:2015ggf}.

An important representative of inhomogeneous spacetimes is the family of the Szekeres universes, which are exact solutions with a pressureless fluid source and no symmetry. This family contains two important cases, a Kantowski-Sachs(-like) inhomogeneous family and a FLRW(-like) inhomogeneous family. More precisely, the diagonal field equations for the scale factors of the Szekeres metric take the form of the anisotropic Kantowski-Sachs cosmological model, or of the isotropic FLRW universe \cite{AndrzejKrasinski:1997zz}. There are various extensions and generalizations of the Szekeres universes in the literature; for example, the Szekeres spacetime with homogeneous fluid source with constant equation of state parameter for the fluid proposed by Szafron in \cite{sz1} leads to the Szekeres-Szafron spacetimes. The introduction of the cosmological constant term in the Szekeres model was studied in \cite{sz2}, while some other extensions can be found in \cite{AndrzejKrasinski:1997zz,sz3,sz5,Barrow:2018zav,Barrow:2019smw,Paliathanasis:2019hvm,Paliathanasis:2020dbe}. In the general case, the metric tensor depends on two scale factors and it has the property that the magnetic part of the Weyl tensor vanishes. Consequently, there is no transfer of information among two cosmic lines through gravitational and sound waves; this is why the Szekeres models are also known as silent universes \cite{Bruni:1994nf}. For more details on the geometric and physical characteristics of the Szekeres spacetimes, we refer the reader to \cite{Goode:1982pg,Apostolopoulos:2016xnm,Apostolopoulos:2016nno,Sussman:2011bp,Georg:2017tpz,Bolejko:2010eb,Mustapha:1999pf,vanElst:1996zs,Maartens:1996uv}.

When the diagonal gravitational field equations of the Szekeres model, together with the continuous equation for the fluid source, are written in the physical variables, they form an algebraic-differential system consisted of four first-order ordinary differential equations supplemented with an algebraic equation. This algebraic-differential system is known as Szekeres system. The conservation laws and the integrability properties of the Szekeres system were investigated in a series of studies \cite{doi:10.1080/14029251.2016.1237199,Sussman:2011bp}. In \cite{Paliathanasis:2017wli}, it was shown that the reduced Szekeres system can be derived from the variational principle of a point-like Lagrangian and the conservation laws can be constructed by the application of Emmy Noether's theorems. This allowed for the exact integration of the classical and quantum system through symmetry considerations, the latter being studied in \cite{Paliathanasis:2018ixu} and it was shown that the quantum dynamics revealed that the time-independent Schr\"{o}dinger equation has a solution in closed-form. The result was that the extrema of the probability density are related to two exact closed-form solutions of the Szekeres system. More interestingly, the probability analysis and the quantum corrections considered by taking into account the Bohmian quantum potential \cite{Bohm:1951xw,Bohm:1951xx} showed that the Szekeres universes remain silent at the semi-classical limit \cite{Paliathanasis:2018ixu}. In \cite{Zampeli:2019jbe}, it was also shown that the second of Hartle's criteria about classicality, i.e. decoherence \cite{Hartle:1986gn} is also satisfied. 

In this work, we extend the previous analysis by considering the existence of a cosmological constant term in the gravitational model. More specifically, we find the linear and quadratic Lie-point symmetries of the reduced system of equations and we study the classical and quantum dynamics. We examine the probability density and the quantum potential term to track the existence of quantum corrections. 

The structure of the paper is the following: in Section 2, we present the Szekeres system with cosmological constant and show that a point-like Lagrangian exists which can describe the reduced system of differential equations. This means that the diagonal field equations follow from a variational principle. Furthermore, by writing the field equations with the use of canonical coordinates we are able to write the additional conservation law which indicates the integrability of the dynamical system. In the context of $H-$normalization the cosmological evolution of the dynamical system is investigated and we find that the unique attractor is an inhomogeneous de Sitter universe. In Section 3, we perform the quantization process and we study the properties of the wavefunction. The results of the quantization are similar with the ones of the system with zero cosmological constant. However, when the ``energy" of the time-independent Schr\"{o}dinger equation vanishes, we are able to construct additional differential invariants. Some of them will lead to solutions which induce quantum corrections at the semi-classical limit. Finally, in Section 4, we draw our conclusions.

\section{Classical dynamical analysis}
In the following we study the classical Szekeres system in the presence of a cosmological constant term. In particular, we investigate the integrability properties of the system by searching for conservation laws and we study the general evolution of the dynamics.
\subsection{General results}
The general spacetime element which describes the Szekeres family is \cite{Szekeres:1974ct}
\begin{equation}
ds^{2}=-dt^{2}+e^{2A(t,x,y,z)}dx^{2}+e^{2B(t,z,y,z)}(dy^{2}+dz^{2})
\end{equation}%
where the functions $A(t,x,y,z),\ B(t,x,y,z)$ are to be specified by the solution of the Einstein equations with energy-momentum tensor of the dust,
\begin{equation}
G_{\mu \nu }=T_{\mu \nu }^{(D)}
\end{equation}%
Instead of the metric variables, we choose the physical variables, since we can take advantage of the fact that in these solutions the two components of
the electric part of the Weyl tensor and the two components of the shear for
the observer denoted by a time-like $4-$vector $u^{\mu }$ are equal
respectively. In these variables, the evolution equations take the form \cite%
{Paliathanasis:2017wli}%
\begin{subequations}
\bal
&\dot{\rho _{t}}+\theta \left( \rho _{t}+p_{t}\right) =0,  \label{sy.01} \\
&\dot{\theta}+\frac{\theta ^{2}}{3}+6\sigma ^{2}+\frac{1}{2}\left( \rho
_{t}+3p_{t}\right) =0,  \label{sy.02} \\
&\dot{\sigma}-\sigma ^{2}+\frac{2}{3}\theta \sigma +E=0,  \label{sy.03} \\
&\dot{E}+3E\sigma +\theta E+\frac{1}{2}\left( \rho _{t}+p_{t}\right) \sigma
=0,  \label{sy.04}
\eal
\end{subequations}
where $\dot{}=u^{\mu }\nabla _{\mu }$, the energy density is defined as $%
\rho =T^{\mu \nu }u_{\mu }u_{\nu }$ and the constraint equation is
\begin{equation}
\frac{\theta ^{2}}{3}-3\sigma ^{2}+\frac{^{\left( 3\right) }R}{2}=\rho _{t},
\end{equation}
For a unified treatment of both the zero and non-zero cosmological constant cases, we have written
\begin{subequations}
\bal
&\rho _{t}=\rho _{D}+\rho _{\Lambda },\\
&p_{t}=p_{D}+p_{\Lambda }
\eal
\end{subequations}
with $\rho _\Lambda=\Lambda$, $p_{D}=0$ and $p_{\Lambda }=-\rho _{\Lambda }$, while we have denoted as $E$ the electric part of the Weyl tensor and as $\rho _{i}$ the density
corresponding to each matter component. Then the system of
equations is written as%
\begin{subequations}
\bal
&\dot{\rho}_{D}+\theta \rho _{D} =0,  \label{ss1.01} \\
&\dot{\theta}+\frac{\theta ^{2}}{3}+6\sigma ^{2}+\frac{1}{2}\left( \rho
_{D}-2\Lambda \right) =0,  \label{ss1.02} \\
&\dot{\sigma}-\sigma ^{2}+\frac{2}{3}\theta \sigma +E =0,  \label{ss1.03} \\
&\dot{E}+3E\sigma +\theta E+\frac{1}{2}\rho _{D}\sigma =0.  \label{ss1.04}
\eal
\end{subequations}
Despite the degeneracy between the different components of the Weyl tensor and the shear, there is no Killing vector field and it can only be said that these solutions are locally axisymmetric \cite{Bruni:1994nf}. However, it is
possible to study this system through the consideration of generalised
symmetries of the differential equations \eqref{sy.01}-\eqref{sy.04}. These
lead to a reduced action and consequently to the possibility of canonically
quantizing the system \cite{Paliathanasis:2018ixu}.

To follow this methodology, we first solve \eqref{ss1.01} and \eqref{ss1.04} with respect to $\theta $ and $\sigma $ respectively 
\begin{subequations}
\bal
&\theta = -\frac{\dot{\rho}_D}{\rho_D},\\
&\sigma = -\frac{\dot{E} - \frac{\dot{\rho}_D}{\rho_D}E}{\frac{1}{2} \rho_D + 3E}
\eal
\end{subequations}
and replace to the other two. Then, we end up with two second-order differential equations which we omit to write. Their transformation to new variables $(x,y)$ is given by 
\begin{subequations}\label{transf}
\bal
&\rho _{D}(x,y) =\frac{6}{y^{2}(( y-x) }, \\
&E(x,y) =-\frac{x}{y^{3}(( y-x) }
\eal
\end{subequations}
with the inverse being 
\begin{subequations}
\bal
&x(\rho _{D},E) =-\frac{6^{4/3}E}{\rho_{D}(6E+\rho _{D})^{1/3}},\\
&y(\rho _{D},E) =\frac{6^{1/3}}{(6E+\rho _{D})^{1/3}}
\eal
\end{subequations}
and leads to the following non-linear system of two second-order differential equations
\begin{subequations}\label{second_order}
\begin{eqnarray}
\ddot{x}-\frac{\Lambda }{3}x-\frac{2x}{y^{3}} &=&0,  \label{s11a} \\
\ddot{y}-\frac{\Lambda }{3}y+\frac{1}{y^{2}} &=&0.  \label{s12b}
\end{eqnarray}
\end{subequations}
We can see that these equations denote integrals of motion since equation \eqref{s12b} comes from $\frac{d}{dt}\left( \dot{y}^{2}-\frac{%
\Lambda }{3}y^{2}-\frac{2}{y}\right)=0 $, that is
\begin{equation}\label{I0}
I_{0}=\dot{y}^{2}-\frac{\Lambda }{3}y^{2}-\frac{2}{y},
\end{equation}%
is a conservation law for the dynamical system and \eqref{s11a} originates from the Hamiltonian of the system, which is also conserved and represents the quantity of energy
\begin{equation}
h=\dot{x}\dot{y}-\frac{\Lambda }{3}xy+\frac{x}{y^{2}}.
\end{equation}
We note here that, contrary to the usual case of gravitational systems, which are constrained due to the presence of the arbitrary functions (the lapse function and the shift vector), the dynamics of this reduced system is not constrained and the Hamiltonian is not a linear combination of the first-class constraints.

Both of these integrals of motion are related to Killing tensors, i.e. they are quadratic in the momenta which are the following:
\begin{subequations}
\be
K^1_{\mu\nu} = 
\begin{pmatrix}
1 & 0 \\
0 & 0%
\end{pmatrix}
\ee
and 
\be
K^2_{\mu\nu} = 
\begin{pmatrix}
0 & 1 \\
1 & 0%
\end{pmatrix}
\ee
\end{subequations}
The conserved quantities are given by $K^i = K^i_{\mu\nu} \dot{q}_\mu \dot{q}_\nu$ and one can see that the Killing tensor $K^1$ corresponds to the integral \eqref{I0}, while $K^2$ to the energy and it is the metric on the configuration space of the dependent variables ${x,y}$. 

It is also not difficult to show that the system of second-order differential equations can be derived
by an action principle with Lagrangian function
\begin{equation}
L\left( x,\dot{x},y,\dot{y}\right) =\dot{x}\dot{y}+\frac{\Lambda }{3}xy-%
\frac{x}{y^{2}}.  \label{reduced_lag}
\end{equation}%
which of course does not have explicit dependence on time, since the Hamiltonian is a conserved quantity. The second-order dynamical system \eqref{s11a}, \eqref{s12b} can be written as a first-order differential system:
\begin{subequations}
\bal
&\dot{x}=u_{x},\\
&\dot{y}=u_{y},\\
&\dot{u}_{x}=\frac{\Lambda }{3}x+\frac{2x}{y},\\
&\dot{u}_{y}=\frac{\Lambda }{3}y-\frac{1}{y^{2}}.
\eal
\end{subequations}
This system admits two stationary points $A=\left( x(A), y(A), u_{x}(A), u_{y}(A)\right)$, they are $A_{1}=\left( 0,\left( \frac{3}{\Lambda }\right) ^{\frac{1}{3}%
},0,0\right)$ and $A_{2}=\left( 0,\left( -\frac{3}{\Lambda }\right) ^{%
\frac{1}{3}},0,0\right) $. The point $A_{1}$ is real when $\Lambda >0$ while
point $A_{2}$ is real for negative cosmological constant. For both stationary points we find that the integrals of motion take the values $h\left( A_{1}\right) =0,h\left( A_{2}\right) =0$
and $I_{0}\left( A_{1,2}\right) =-3^{\frac{2}{3}}\left\vert \Lambda
\right\vert ^{\frac{1}{3}}$, which leads to the conclusion that, in order to the stationary points exist, the condition $I_{0}<0$ must be satisfied. In addition to that, the stationary points are sources, because at least one of the eigenvalues of the linearized system is always positive.

In the limit that both constants of motion vanish, $h=0,I_{0}=0$, the system acquires a solution in closed form
\begin{subequations}
\bal
&x=x_{0}\exp \left( -\Xi \left( t\right) \right) ,\\
&y=\frac{\sqrt[3]{6}\left( \sinh \left( \frac{1}{2}\sqrt{3\Lambda }\left(
t-t_{0}\right) \right) \right) {}^{2/3}}{\sqrt[3]{\Lambda }}.
\eal
\end{subequations}
where 
\small{
\begin{equation}
\Xi \left( t\right) =\frac{\sinh \left( \sqrt{3\Lambda }\left(
t-t_{0}\right) \right) \left( 3\log \left( \cosh \left( \frac{1}{2}\sqrt{%
3\Lambda }\left( t-t_{0}\right) \right) \right) -\log \left( \sinh \left(
\frac{1}{2}\sqrt{3\Lambda }\left( t-t_{0}\right) \right) \right) \right) }{3%
\sqrt{2}\sqrt{\Lambda }\left( \frac{\left( -\sinh \left( \frac{1}{2}\sqrt{%
3\Lambda }\left( t-t_{0}\right) \right) \right) {}^{2/3}}{\sqrt[3]{\Lambda }}%
\right) {}^{3/2}\sqrt{\cosh \left( \sqrt{3\Lambda }\left( t-t_{0}\right)
\right) +1}}
\end{equation}}
and $x_{0}\,\ $and $t_{0}$ are constants of integrations. In this case, the curvature also simplifies as
\begin{equation}
R=\frac{1}{8}\left( 15\Lambda +2\sqrt{3}\sigma \sqrt{\Lambda +6\rho
_{D}^{3}E+\rho _{D}^{4}}-6\rho _{D}^{3}E-\rho _{D}^{4}+16\rho _{D}+45\sigma
^{2}\right)
\end{equation}
In conclusion, the four-dimensional Szekeres spacetime in the presence of a cosmological constant term is an integrable system. This result extends the previous analysis on the integrability of the original Szekeres system \cite{Paliathanasis:2017wli}.
\subsection{Stability analysis}
In order to understand the dynamics and the general evolution of the Szekeres system with the cosmological constant \eqref{ss1.01}-\eqref{ss1.04}, we now define the new dimensionless variables
\begin{equation}
\Omega _{m}=\frac{3\rho }{\theta ^{2}}, \ \Omega _{\Lambda }=\frac{\Lambda }{\theta ^{2}}, \ \Sigma =\frac{\sigma }{\theta }, \ \alpha =\frac{E}{\theta ^{2}}, \ \Omega _{R}=\frac{3}{2}\frac{^{\left( 3\right) }R}{\theta ^{2}}.
\label{dd.01}
\end{equation}%
In this case, the Szekeres system is written as
\begin{subequations}
\bal
&\Omega _{m}^{\prime } =\frac{1}{3}\Omega _{m}\left( 36\Sigma ^{2}+\Omega
_{m}-6\Omega _{\Lambda }-1\right) ,  \label{dd.02} \\
&\Sigma ^{\prime } =-\alpha +\frac{1}{6}\Sigma \left( 6\Sigma \left(
1+6\Sigma \right) -2+\Omega _{m}-6\Omega _{\Lambda }\right) ,  \label{dd.03}
\\
&\alpha ^{\prime } =\frac{1}{6}\left( -\Sigma \Omega _{m}+2\alpha \left(
9\Sigma \left( 4\Sigma -1\right) +\Omega _{m}-6\Omega _{\Lambda }-1\right)
\right) ,  \label{dd.04} \\
&\Omega _{\Lambda }^{\prime } =\frac{1}{3}\Omega _{\Lambda }\left( 36\Sigma
^{2}+\Omega _{m}-6\Omega _{\Lambda }+2\right) ,  \label{dd.05}
\eal
\end{subequations}
with algebraic equation%
\begin{equation}
\Omega _{R}=9\Sigma ^{2}+\Omega _{m}+3\Omega _{\Lambda }-1,  \label{dd.06}
\end{equation}
and the prime $'$ denotes total derivative with respect to the new independent variable $d\tau =\theta dt$. Every stationary point $P=\left( \Omega _{m}\left( P\right) ,\Sigma \left(
P\right) ,\alpha \left( P\right) ,\Omega _{\Lambda }\left( P\right) \right) $
of the dynamical system \eqref{dd.02}-\eqref{dd.05} corresponds to an exact
solution. The calculation of the dynamical system and the study of their
stability are essential in order to understand the various eras provided
by the model as well as to understand the evolution of the variables in the phase space. The dynamical system consisted by the equations \eqref{dd.02}-\eqref{dd.05} admits nine physically accepted stationary points, six of which correspond to the original Szekeres system \cite{Bruni:1994nf} with zero cosmological constant:
\begin{itemize}
\item Point $P_{1}=( 0,0,0,0)$, with $\Omega _{R}=-1$ describes the
isotropic Milne universe. The eigenvalues of the linearized system around
the stationary point $P_{1}$ are $e_{1}( P_{1})=\frac{2}{3}%
,e_{2}( P_{1})=-\frac{1}{3},e_{3}( P_{1}) =-\frac{1}{3}$ and $e_{4}(P_{1}) =-\frac{1}{3}$ from where we conclude
that $P_{1}$ is a saddle point and the (inhomogeneous) Milne universe is
always an unstable solution.

\item Point $P_{2}=( 1,0,0,0) $ with $\Omega _{R}( P_{2}) =0$
describes a spatially flat FLRW (-like) universe dominated by the dust fluid
source. The eigenvalues are derived $e_{1}( P_{2})
=1,e_{2}( P_{2}) =-\frac{1}{2},e_{3}( P_{2}) =\frac{1%
}{3}$ and $e_{4}( P_{2}) =\frac{1}{3},$ which means that $P_{2}$
is always a saddle point.

\item Stationary points $P_{3}=( 0,-\frac{1}{3},0,0) $ and $%
P_{4}=( 0,\frac{1}{3},\frac{2}{9},0) $ with eigenvalues $%
e_{1}( P_{3}) =2,e_{2}( P_{3}) =2,e_{3}(
P_{3}) =1$, $e_{4}( P_{3}) =1$ and $e_{1}(
P_{4}) =2,e_{2}( P_{4}) =\frac{5}{3},e_{3}(
P_{4}) =1$ and $e_{4}( P_{4}) =\frac{2}{3}$; describe
unstable Kasner (-like) universes, i.e $\Omega _{R}( P_{3}) =0$, $%
\Omega _{R}( P_{4}) =0$. In particular, points $P_{3}$ and $P_{4}$
are sources.

\item Kantowski-Sachs (-like) universes are provided by the exact solutions at the stationary points $P_{5}=( 0,\frac{1}{6},0,0) $ and $P_{6}=(
0,-\frac{1}{12},\frac{1}{32},0) $ with $\Omega _{R}( P_{5})
=-\frac{3}{4}$and $\Omega _{R}( P_{6}) =-\frac{15}{16}$. The
eigenvalues of the linearized systems around the stationary points are $%
e_{1}( P_{5}) =1,e_{2}( P_{5}) =-\frac{1}{2}%
,e_{3}( P_{5}) =\frac{1}{2}$, $e_{4}( P_{5}) =0$ and
$e_{1}( P_{6}) =\frac{3}{4},e_{2}( P_{6}) =-\frac{5}{8%
},e_{3}( P_{6}) =\frac{1}{4}$ and $e_{4}( P_{6}) =-%
\frac{1}{4}$; from where we conclude that the stationary points $P_{5}$ and $%
P_{6}$ are saddle points, that is, the Kantowksi-Sachs (-like) universes are
unstable solutions.
\end{itemize}
The following three points appear when $\Omega _{\Lambda }\neq 0$:
\begin{itemize}
\item Points $P_{7}=( 0,-\frac{1}{3},\frac{1}{3},1) $ and $P_{8}=(
0,\frac{2}{3},0,3) $ provide $\Omega _{R}( P_{7})
=3,\Omega _{R}( P_{8}) =12$, which means that the exact
solutions at the points describe Bianchi III universes. The eigenvalues are $%
e_{1}( P_{7}) =1,e_{2}( P_{7}) =-1,e_{3}(
P_{7}) =-1$, $e_{4}( P_{7}) =-2$ and $e_{1}(
P_{8}) =2,e_{2}( P_{8}) =-1,e_{3}( P_{8}) =-2$
and $e_{4}( P_{8}) =-3$, which means that the stationary points $%
P_{7},P_{8}$ are sources and the exact solutions are always unstable.

\item Finally, point $P_{9}=( 0,0,0,1) $ describe is an attractor since
all the eigenvalues are negative, that is, $e_{1}( P_{9})
=-1,e_{2}( P_{9}) =-1,e_{3}( P_{9}) =-\frac{2}{3}$, $%
e_{4}( P_{9}) =-\frac{2}{3}$, and describe the de Sitter (-like)
universe, while $\Omega _{R}( P_{9}) =0$.
\end{itemize}
In addition to the above-mentioned, physically accepted stationary points, the dynamical system \eqref{dd.02}-\eqref{dd.05} admits a stationary point $\bar{P}=( -3,-\frac{1}{3},\frac{1}{6},0)$ which is not physically accepted. It is important to mention that, in the presence of the cosmological constant, the only attractor is the de Sitter point $P_{9}$, indicating that the existence of a positive cosmological constant in the dynamical system leads to the isotropization of the Szekeres universe. Note that, with no cosmological constant, the attractors of the Szekeres system are the Kantowski-Sachs spacetimes \cite{Bruni:1994nf}.
\section{Quantization}
In this Section we study the quantization of the Szekeres system with the cosmological constant term.
\subsection{Quadratic integrals}
We are now interested in studying the quantum dynamics of the physical system, in order to examine the existence of quantum corrections at the semi-classical limit. The approach we follow involves the consideration of the classical constants of motion at the quantum level, by imposing them on the wavefunction. This leads to a Schr\"odinger-like dynamical equation, while the additional symmetry corresponds to a non-homogeneous quadratic in momenta integral of the reduced Lagrangian \eqref{reduced_lag}. Since the dynamics is not constrained, the analysis is simplified and it is possible to define a probability density and have a notion of time in this reduced configuration space. In order to quantize the system, we consider the quantum form of the quadratic symmetries $h, I_0$ of the classical reduced action \eqref{reduced_lag} imposed on the wavefunction. As operators acting on the Hilbert space, they take the general form
\bal\label{quantumoperators}
&\hat{H} = -\frac{1}{\mu} \partial_\mu \left( \mu K_2^{\mu\nu} \partial_\beta\right)+V(q^i), \\
&\hat{K}_1  =\frac{1}{\mu}\partial_\mu \left( \mu K_1^{\mu\nu} \partial_\nu \right)
\eal
where $\mu= \sqrt{\abs{K^2_{\mu\nu}}}$ is the measure on the phase space. For the specific problem, the equations \eqref{quantumoperators} take the explicit form 
\bes\label{quadratic_quantum}
\begin{eqnarray}
\left( -\partial _{xx}-\frac{2}{y}-\frac{\Lambda y^{2}}{3}\right) \Psi
&=&I_{0}\Psi ,  \label{wdw} \\
\left( -\partial _{xy}+\frac{x}{y^{2}}-\frac{\Lambda }{3}xy\right) \Psi
&=&h\Psi ,  \label{wdw2}
\end{eqnarray}
\ees
The general solution of the system \eqref{quadratic_quantum} is
\begin{equation}
\Psi (x,y;h,I_{0})=A(y)\left( \Psi _{1}\cos (\frac{x}{A(y)}+h\int A\left(
y\right) )+\Psi _{2}\sin (\frac{x}{A(y)}+h\int A\left( y\right) )\right)
\label{wavefunction}
\end{equation}%
where $A(y)=\frac{\sqrt{3y}}{\sqrt{6+3I_{0}y+\Lambda y^{3}}}$. We consider the special cases in which at least one of the eigenvalues $h, I_{0}$ of the quantum operators vanish. As a consequence, the form of the solution to the wavefunction is significantly simplified. Indeed, for $h=0$, the wavefunction becomes
\begin{equation}
\Psi _{A}\left( x,y;0,I_{0}\right) =\frac{\sqrt{y}}{\sqrt{6+I_{0}y+\Lambda
y^{3}}}\left( \Psi _{1}\cos \frac{x}{A(y)}+\Psi _{2}\sin \frac{x}{A(y)}%
\right)
\end{equation}%
while when $I_{0}=0$ it takes the form
\begin{equation}
\Psi _{B}\left( x,y;h,0\right) =\frac{\sqrt{y}}{\sqrt{6+\Lambda y^{3}}}%
\left( \Psi _{1}\cos C\left( x,y\right) +\Psi _{2}\sin C(x,y)\right)
\end{equation}%
with
\be
C(x,y)=\frac{2h_{0}\sinh ^{-1}\left( \frac{\sqrt{\Lambda }y^{3/2}}{\sqrt{6}}\right) }{\sqrt{3}\sqrt{\Lambda }}+x\sqrt{\frac{6+\Lambda y^{3}}{3y}}.
\ee 
Finally, when both constants vanish, $h=0$ and $I_{0}=0$, the wavefunction acquires the form
\begin{equation}
\Psi _{C}\left( x,y;0,0\right) =\sqrt{\frac{y}{6+\Lambda y^{3}}}\left( \Psi
_{1}\cos \left( x\sqrt{\frac{6+\Lambda y^{3}}{3y}}\right) +\Psi _{2}\sin
\left( x\sqrt{\frac{6+\Lambda y^{3}}{3y}}\right) \right)
\end{equation}
We focus on the special case in which $h=0$, i.e. the energy of the system vanishes. The wavefunction is described by
the functional form of $\Psi _{A}$ for a general value of $I_0$ and $\Psi _{C}$ for $I_0$. For these two cases, the extrema of the probability density $P( \Psi )
=\vert \Psi \vert ^{2}$ for $\Psi _{2}=0$ are calculated on the
points with coordinates $\{ x=\pi \frac{N}{2}\sqrt{\frac{3y}{6+\Lambda
y^{3}}},\Lambda y^{3}-3=0,N=2n,n\in \mathbb{N} \}$ and $\{ x=\pi \frac{N}{2}\sqrt{\frac{3y}{6+\Lambda y^{3}}}%
,y,N=( 2n+1) ,n\in\mathbb{N} \}$. On the other hand, for $\Psi _{1}=0$, the stationary points are on the points $\{ x=\pi \frac{N}{4}\sqrt{\frac{3y}{6+\Lambda y^{3}}}, \Lambda y^{3}-3=0,N=2n, n\in \mathbb{N}\}$ and $\{ x=\pi \frac{N}{4}\sqrt{\frac{3y}{6+\Lambda y^{3}}},y,N=( 2n+1), n\in\mathbb{N}\}$. At these points of the configuration space, observations of the correlations between observables must be precluded according to Hartle's criterion \cite{Hartle:1986gn}.

\begin{figure}[ht]\centering
\includegraphics[width=0.5\textwidth]{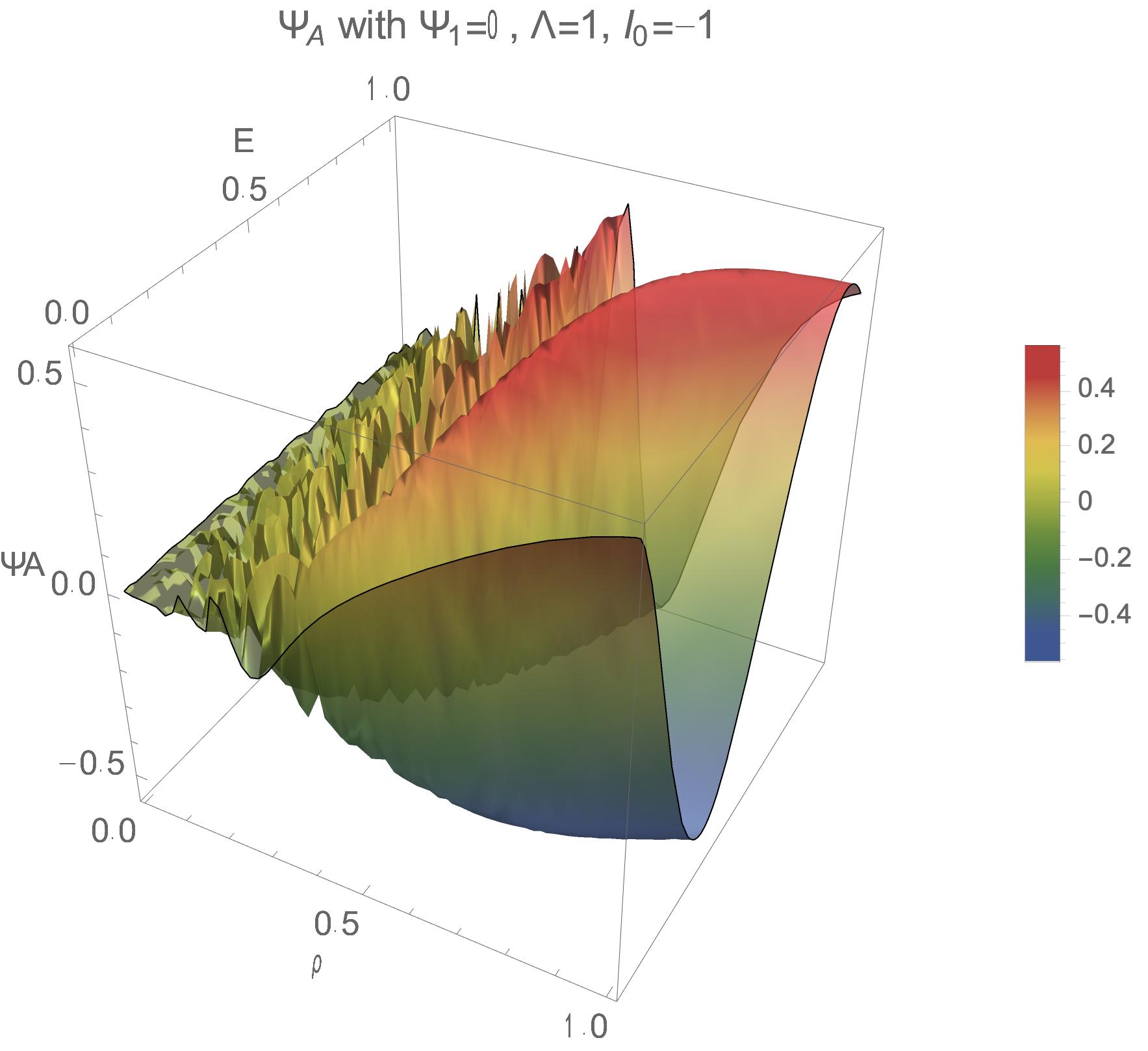}\hfil
\includegraphics[width=0.5\textwidth]{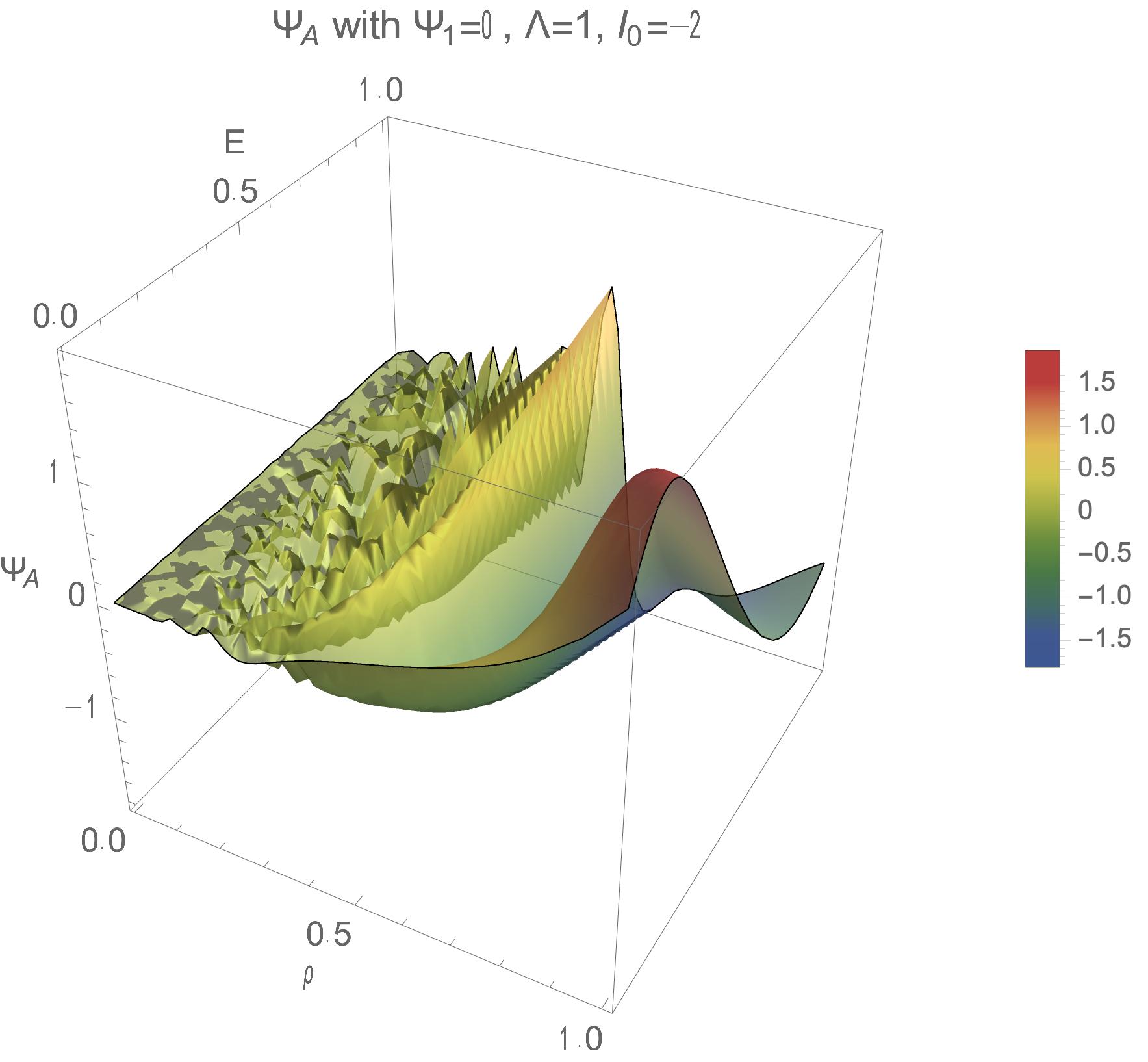}\par\medskip
\includegraphics[width=0.5\textwidth]{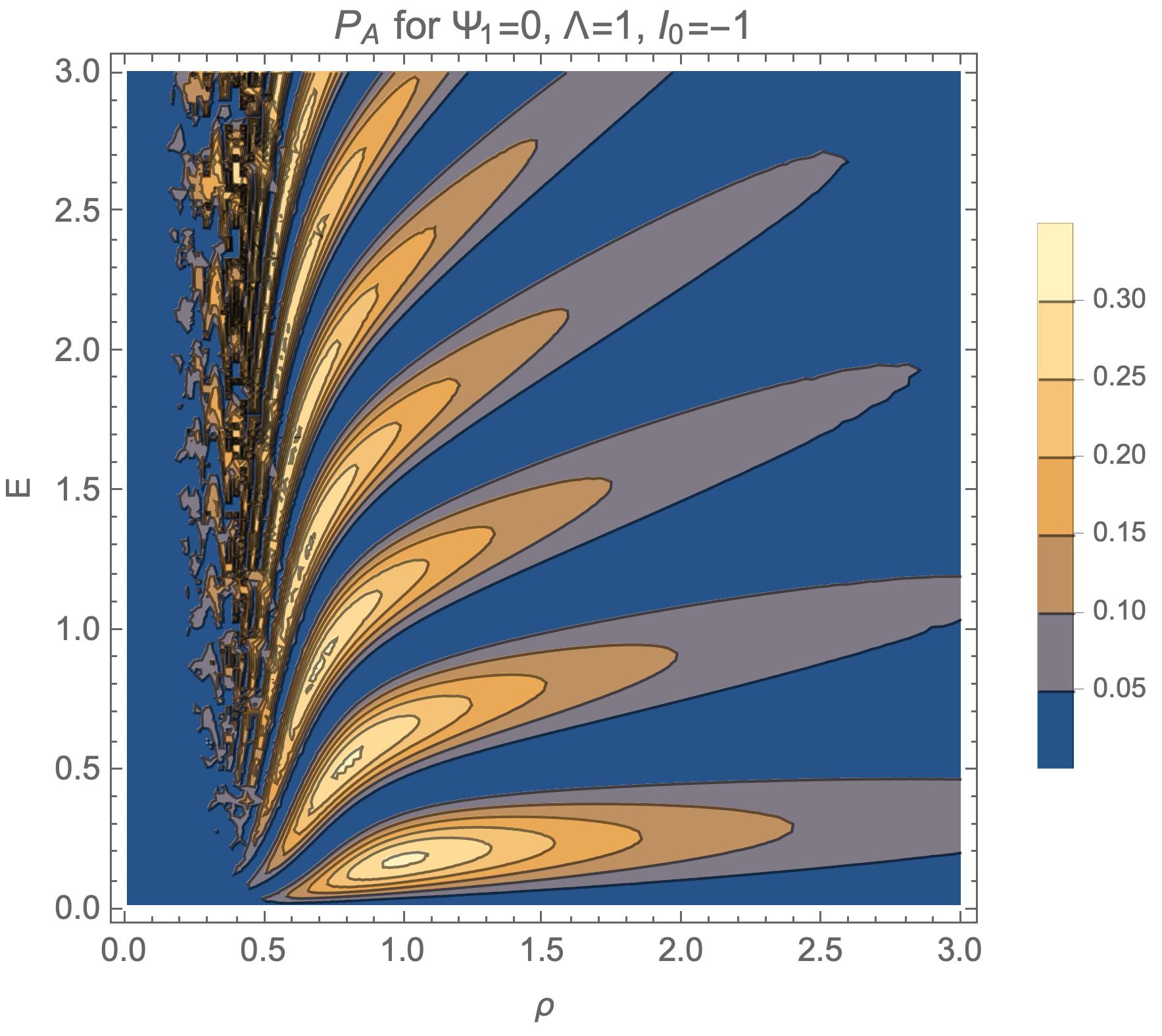}\hfil
\includegraphics[width=0.5\textwidth]{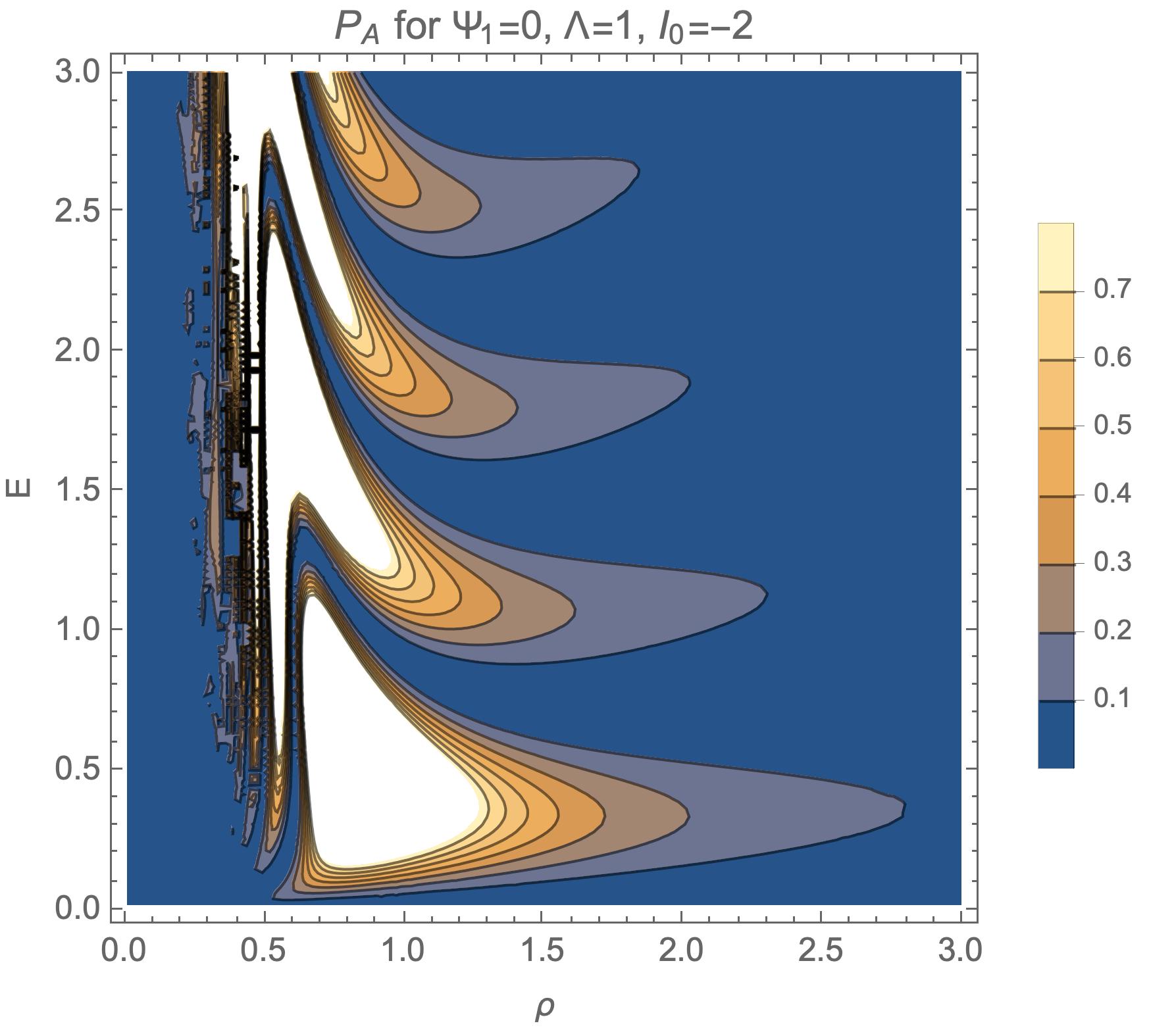}
\caption{Qualitative evolution of the wavefunction $\Psi _{A}$ and of the
probability density $P_{A}=\left\vert \Psi _{A}\right\vert ^{2}$ for $\Psi _{1}=0$, $\Psi _{2}=1$, $\Lambda =1$ and for $I_{0}=-1$ (left column) and $I_{0}=-2$ (right column). The plots are presented in the original variables $%
\left\{ \protect\rho, E\right\}$.}
\label{fig1}
\end{figure}

\begin{figure}[ht]\centering
\includegraphics[width=0.5\textwidth]{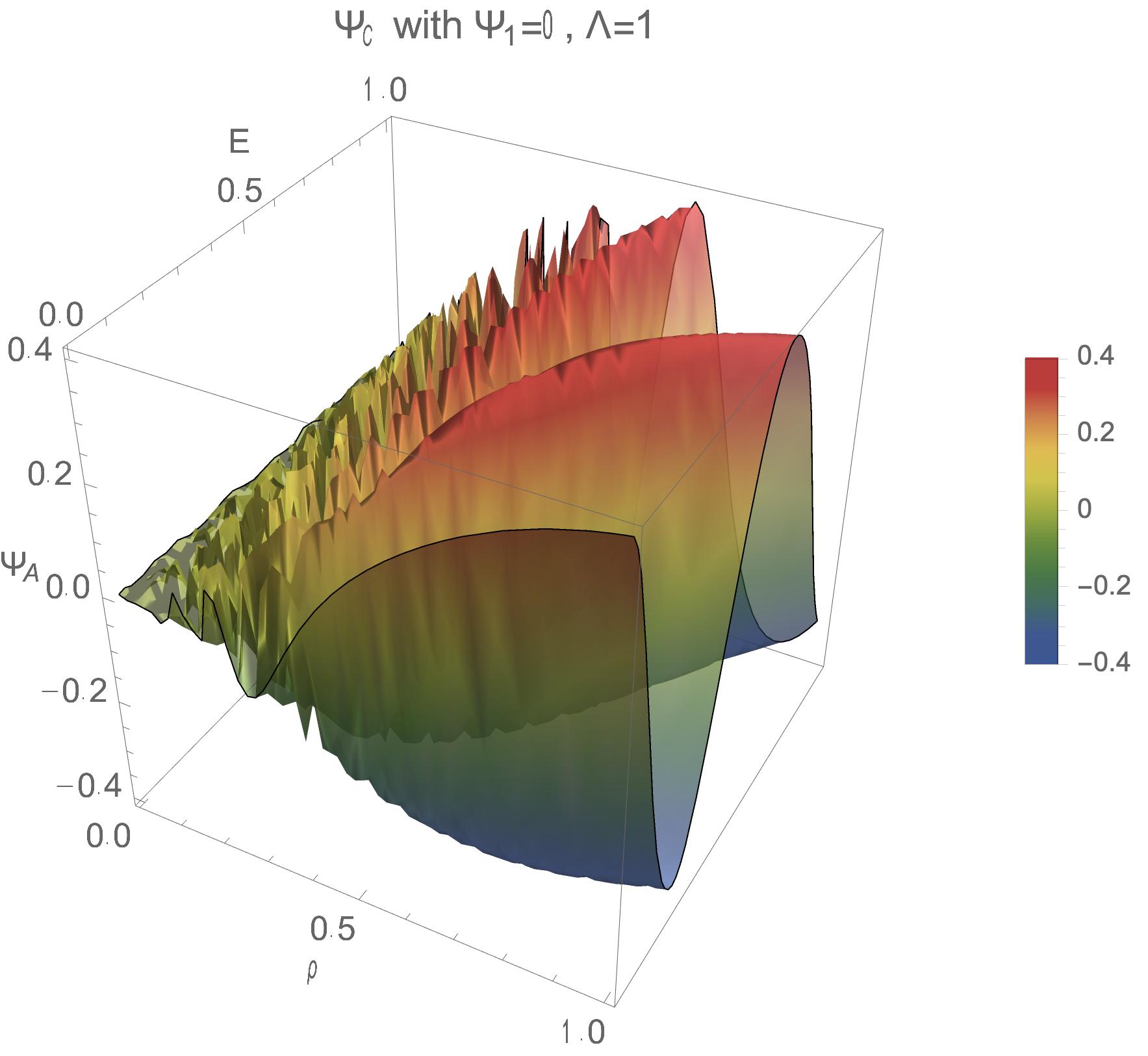}\hfil
\includegraphics[width=0.5\textwidth]{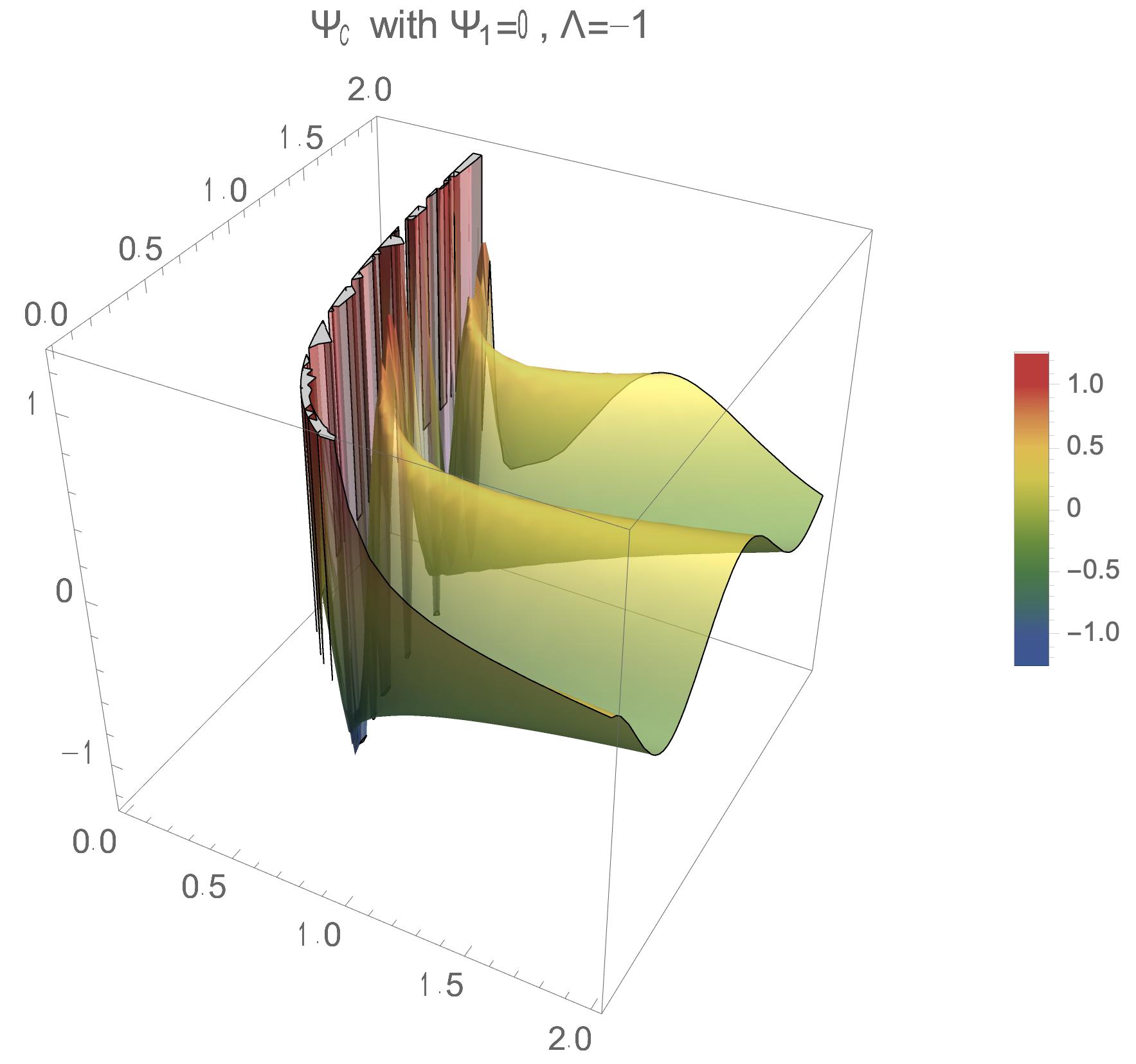}\par\medskip
\includegraphics[width=0.5\textwidth]{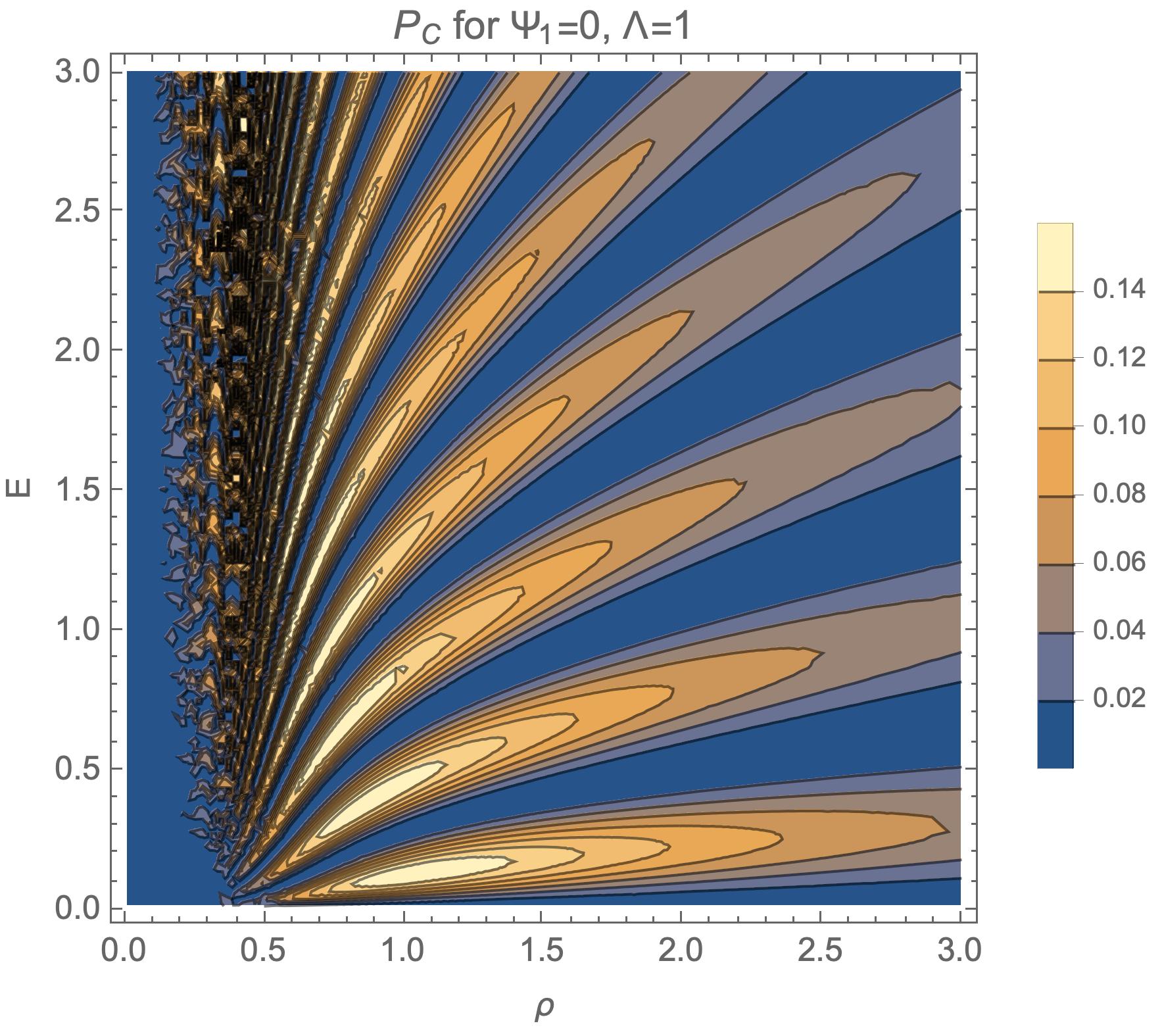}\hfil
\includegraphics[width=0.5\textwidth]{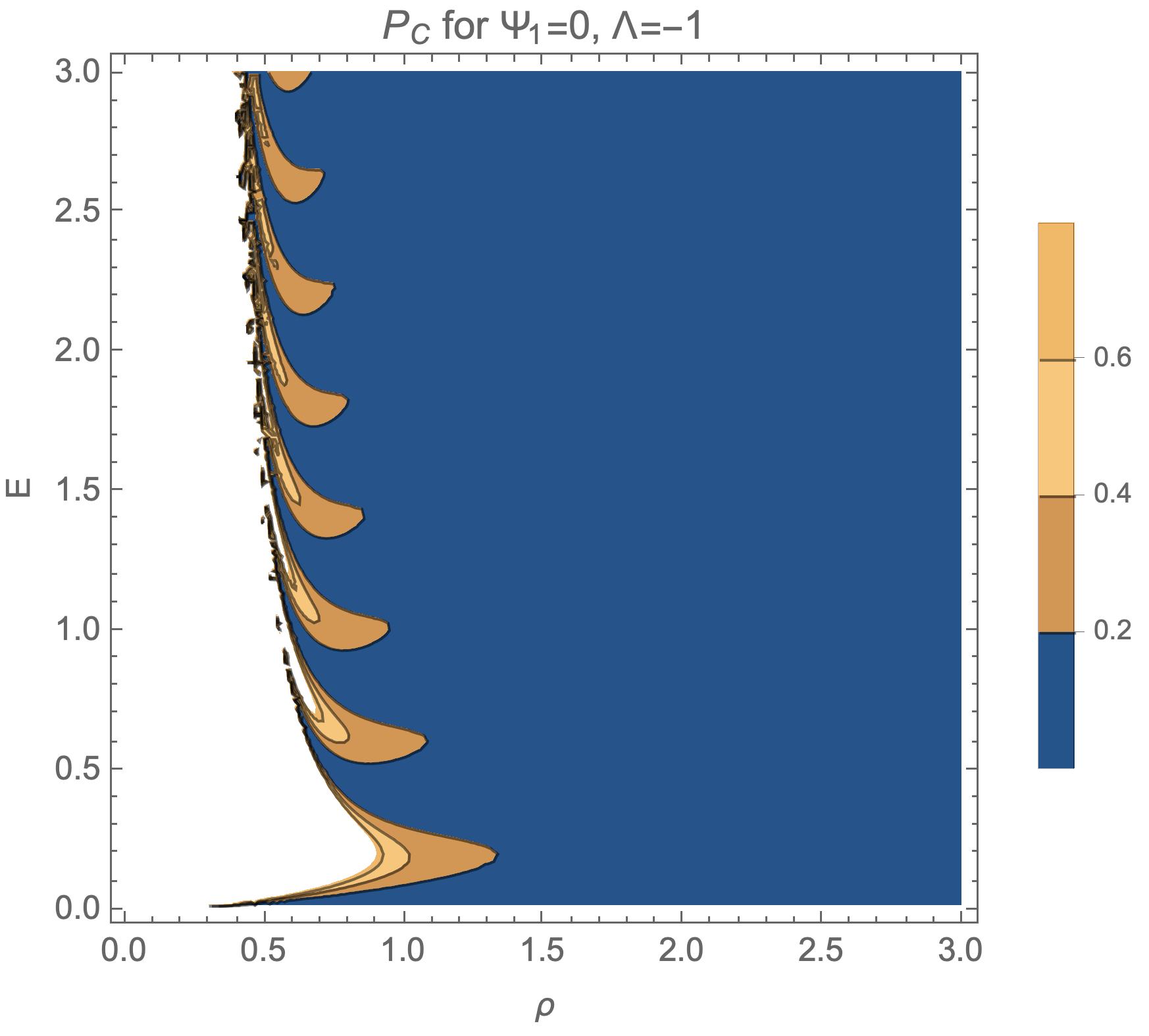}
\caption{Qualitative evolution of the wavefunction $\Psi _{C}$ and of the
probability density $P_{C}=\left\vert \Psi _{C}\right\vert ^{2}$ for $\Psi _{1}=0$,$\Psi _{2}=1$ and for $\Lambda =1$ (left column) and $\Lambda
=-1 $ (right column). The plots are presented in the original variables $%
\left\{ \protect\rho, E\right\} .$}
\label{fig2}
\end{figure}

For the wavefunction $\Psi _{A}$ and $\Psi _{1}=0$, we find that%
\begin{equation}
\Psi _{A}\left( \rho ,E\right) \simeq -\Psi _{2}\frac{E^{\frac{1}{3}}\rho }{%
\sqrt{\Lambda }}\sin \left( 2\sqrt{3\Lambda }\frac{E^{\frac{1}{3}}}{\rho }%
\right)
\end{equation}%
with limit $\lim_{\rho \rightarrow 0}\Psi _{A}\left( \rho ,E\right) =0$. This region must also be precluded. These results for $\Psi _{A},\Psi _{C}$ and the corresponding probability densities $P\left( \Psi _{A}\right) ,P\left( \Psi _{C}\right) $ results appear in Figs \ref{fig1} and \ref{fig2}.

Another way to verify the above analysis is to calculate the quantum potential of Bohmian mechanics, in the de Broglie-Bohm approximation \cite{de_Broglie_1927,holland1995quantum,bohm2006undivided} which is given by the relation
\be\label{quantumpot}
\mathcal{Q} (q^i) = -\frac{\square \Omega (q^i)}{2 \Omega (q^i)}. 
\ee
where $\Omega (q^i)$ is the amplitude of the wavefunction in polar form, $\Psi(q^i) = \Omega(q^i) e^{i S(q^i)}$, $q^i$ are the variables of the configuration space, in this case $(x,y)$ and $\square$ is the Laplacian for this space. This appears as an additional term in the Hamilton-Jacobi equation 
\begin{equation}\label{modified_HJ} 
\frac{1}{2} K_2^{ij} (q) \frac{\partial S}{\partial q^i} \frac{\partial S}{\partial q^j} -\mathcal{Q} (q) +V (q)=0,
\end{equation}
in which $S$ is defined through the semiclassical equations 
\be\label{semiclass}
\frac{\partial S}{\partial q^i}=p_i =\frac{\partial L}{\partial \dot{q}^i}.
\ee
Whether the quantum potential vanishes or not signifies the existence or not respectively of quantum deviations of the paths. There is also a second equation which appears when one substitutes the polar form in the Schr\"odinger equation which is interpreted as the continuity equation 
\begin{equation}
K_2^{ij} \partial_i S \partial_j \Omega + \frac{\Omega}{ 2 \mu} \partial_i (\mu K_2^{ij} \partial_j S)=0.
\end{equation}
Substituting $\Omega$ from our solution in \eqref{quantumpot}, we find that the quantum potential indeed vanishes, thus indicating that this wavefunction leads to the classical solution. This also means that the paths on the configuration space are the classical ones. Therefore, this approach agrees with the previous results.
\subsection{Linear integrals quantization}
We now consider the Schr\"{o}dinger equation \eqref{wdw2} with $h=0$. In this special case, we are interested in the existence of linear symmetries and the corresponding linear integrals of motion. In order to find them, we investigate the one-parameter point transformations which leave the equation invariant, that is, we search for Lie point symmetries \cite{stephani1989differential,Paliathanasis:2013tza,Christodoulakis:2013xha,Terzis:2015mua,Tsamparlis:2018nyo}.

The Lie point symmetries of \eqref{wdw2} with $h=0$ are the following
vector fields%
\begin{subequations}
\begin{align}
&X_{1}=\frac{1}{x}\partial _{x},\\
&X_{2}=\frac{y^{2}}{\Lambda y^{3}-3}\partial
_{y},\\
&X_{3}=x\partial _{x}-\frac{y\left( \Lambda y^{3}+6\right) }{\Lambda y^{3}-3}\partial _{y},\\
&X_4=\Psi \partial _{\Psi }
\end{align}
\end{subequations}
plus the vector field 
\be
X_{\beta }=\beta \left( x,y\right) \partial _{\Psi }
\ee
where $\beta \left( x,y\right) $ is a solution of the original equation
and indicates the infinite number of solutions. Moreover, the set of the
vector fields $\{ X_{4},X_{\beta}\}$ is directly related
to the linearity of the differential equation. In tables \ref{tab1}, \ref{tab2} we present the commutators and the adjoint representation for the admitted finite Lie algebra.


\begin{table}[ht] \centering
\caption{Commutator table for the Lie point symmetries for the
time-independent Schr\"{o}dinger equation \eqref{wdw2}.}
\begin{tabular}{ccccc}
\hline\hline
$\left[ X_{I},X_{J}\right] $ & $\mathbf{X}_{1}$ & $\mathbf{X}_{2}$ & $%
\mathbf{X}_{3}$ & $\mathbf{X}_{4}$ \\ \hline
$\mathbf{X}_{1}$ & $0$ & $0$ & $2X_{1}$ & $0$ \\
$\mathbf{X}_{2}$ & $0$ & $0$ & $-2X_{2}$ & $0$ \\
$\mathbf{X}_{3}$ & $-2X_{1}$ & $-2X_{2}$ & $0$ & \thinspace $0$ \\
$\mathbf{X}_{4}$ & $0$ & $0$ & $0$ & $0$ \\ \hline\hline
\end{tabular}%
\label{tab1}%
\end{table}%

\begin{table}[ht] \centering%
\caption{Adjoint representation for the Lie point symmetries for
the time-independent Schr\"{o}dinger equation \eqref{wdw2}.}%
\begin{tabular}{ccccc}
\hline\hline
$Ad\left( e^{ \varepsilon \mathbf{X}_{i}}\right) \mathbf{X}%
_{j} $ & $\mathbf{X}_{1}$ & $\mathbf{X}_{2}$ & $\mathbf{X}_{3}$ & $\mathbf{X}%
_{4}$ \\ \hline
$\mathbf{X}_{1}$ & $X_{1}$ & $X_{2}$ & $-2\varepsilon X_{1}+X_{3}$ & $X_{4}$
\\
$\mathbf{X}_{2}$ & $X_{1}$ & $X_{2}$ & $2\varepsilon X_{2}+X_{3}$ & $X_{4}$
\\
$\mathbf{X}_{3}$ & $e^{2\varepsilon }X_{1}$ & $e^{-2\varepsilon }X_{2}$ & $%
X_{3}$ & $X_{4}$ \\
$\mathbf{X}_{4}$ & $X_{1}$ & $X_{2}$ & $X_{3}$ & $X_{4}$ \\ \hline\hline
\end{tabular}%
\label{tab2}%
\end{table}

The Lie point symmetries $X_i$ of the equation \eqref{wdw2} define the symmetry group of the system under consideration. When we take into account this symmetry group at the quantum level, we consider only the commutative subalgebras to ensure locality. When the quantum operators related to these symmetries act on the wavefunction, they leave the solution $\Psi$ independent of the variable related to the symmetry, e.g. if $\hat{Q}_i$ is an operator which is related to the variable $\Lambda_i$, then
\be\label{eigenvalueeqs}
\hat{Q}_i \Psi (q,\Lambda_i) \equiv -\frac{i}{2 \mu} (\mu X_i^\alpha \partial_\alpha +\partial_\alpha \mu X_i^\alpha) \Psi (q, \Lambda_i) \Rightarrow \Psi (q, \Lambda_i) \equiv \Psi (q)
\ee
We also note that these Lie-point symmetries are related to the variational symmetries of the reduced Lagrangian \cite{Christodoulakis:2013xha}.

From the Tables \ref{tab1}, \ref{tab2}, we can determine all the possible one-dimensional Lie algebras which can provide independent constraint equations. This classification is known as the one-dimensional system. It follows that the independent one-dimensional Lie algebras
are: $\{ X_{1}\}$, $\{ X_{2}\}$, $\{ X_{3}\}$, $\{ X_{4}\}$, $\{ X_{1}-\alpha X_{2}\}$, $\{
X_{1}+i\gamma X_{4}\}$, $\{ X_{2}+i\gamma X_{4}\}$, $\{
X_{3}+i\gamma X_{4}\}$ and $\{ X_{1}+\alpha X_{2}+\gamma
X_{4}\}$.
These one-dimensional subalgebras give us the following quantum solutions:
\begin{itemize}
\item From the quantum equations provided by the vector fields $\left\{ X_{1}\right\}$, $\left\{ X_{2}\right\}$, we find the trivial solution $\Psi \left(x,y\right) =0$, thus we omit it. In this case, the probability and the Bohmian analysis coincide and according to Hartle's criterion \cite{Hartle:1986gn}, correlations in this region of the configuration space are precluded.

\item On the other hand, the vector field $X_{3}$ provides the differential operator 
\begin{equation}
\left( x\partial _{x}-\frac{y\left( \Lambda y^{3}+6\right) }{\Lambda y^{3}-3}%
\partial _{y}\right) \Psi\left( x,y\right) =0,
\end{equation}%
which has the solution 
\begin{equation}
\Psi _{3}\left( x,y\right) =\Psi _{1}^{0}J_{0}\left( \frac{x\sqrt{\Lambda
y^{2}+6}}{\sqrt{3y}}\right) +\Psi _{2}^{0}Y_{0}\left( \frac{x\sqrt{\Lambda
y^{2}+6}}{\sqrt{3y}}\right) ,  \label{ww1}
\end{equation}%
where $J_{0}\left( \mathbf{x}\right) ,Y_{0}\left( \mathbf{x}\right) $ are
the Bessel functions of the first kind. We remark that the wavefunction $\Psi _{3}\left( x,y\right)$ does not satisfy eq. \eqref{wdw}, thus indicating itself as a new solution.
\item The linear combination $\left\{ X_{1}-aX_{2}\right\} $ gives the following solution for the wavefunction
\begin{equation}
\Psi _{4}\left( x,y\right) =\Psi _{1}^{0}\cos \left( \frac{\sqrt{3}\left(
ayx^{2}+\Lambda y^{3}+6\right) }{6\sqrt{a}y}\right) +\Psi _{2}^{0}\cos
\left( \frac{\sqrt{3}\left( ayx^{2}+\Lambda y^{3}+6\right) }{6\sqrt{a}y}%
\right) ,
\end{equation}%
which is also a new solution.

\item Similarly from $\left\{ X_{1}-\alpha X_{2}+i\gamma X_{4}\right\}$, it
follows
\begin{equation}
\Psi _{124}\left( x,y\right) =e^{\frac{\gamma }{2}x^{2}}\left( \Psi
_{1}^{0} e^{ i\kappa _{+}\left( x,y\right)} +\Psi _{2}^{0}e^{i\kappa _{-}\left( x,y\right)} \right)
\end{equation}
where $\kappa _{\pm }\left( x,y\right) =-\left( 3\gamma \pm \sqrt{9\gamma
^{2}+12a}\right) \frac{\left( ayx^{2}+\Lambda y^{3}+6\right) }{2\sqrt{a}y}$.

\item From $\left\{ X_{1}+i\gamma X_{4}\right\}$ and $\left\{
X_{2}+i\gamma X_{4}\right\} $, we calculate the analytic solutions%
\bal
&\Psi _{14}\left( x,y\right) =\Psi _{1}^{0}\exp \left( \frac{i}{6\gamma y}%
\left( \Lambda y^{3}+6+3x^{2}y^{2}\gamma ^{2}\right) \right) ,\\
&\Psi _{24}\left( x,y\right) =\Psi _{1}^{0}\exp \left( \frac{i}{6\gamma y}%
\left( 3\gamma ^{2}\left( \Lambda y^{3}+6\right) +3x^{2}y^{2}\right) \right).
\eal

\item Finally, from $\left\{ X_{3}+i\gamma X_{4}\right\} $, it follows the known solution $\Psi _{C}\left( x,y\right)$, which satisfies the constraint equation \eqref{wdw}.
\end{itemize}
Now, if we use \eqref{ww1} to calculate the quantum potential of these wavefunctions, we find that in all cases except one, it vanishes. In particular, for the case $\Psi _{3}\left( x,y\right)$ we find that, at the limit $\frac{x\sqrt{\Lambda y^{2}+6}}{\sqrt{3y}}\rightarrow \infty$, the wavefunction is approximated as
\small{
\begin{equation}
\Psi _{3}\left( x,y\right) =\sqrt{\frac{2}{\pi }\sqrt{\frac{3y}{x^{2}\left(
\Lambda y^{2}+6\right) }}}\left( \Psi _{1}^{0}\cos \left( \frac{x\sqrt{%
\Lambda y^{2}+6}}{\sqrt{3y}}-\frac{\pi }{4}\right) +\Psi _{2}\sin \left(
\frac{x\sqrt{\Lambda y^{2}+6}}{\sqrt{3y}}-\frac{\pi }{4}\right) \right)
\end{equation}}
In this case the quantum potential takes the non-zero form 
\begin{equation}
\mathcal{Q}\left( x,y\right) = \frac{\Lambda y^{3}-3}{8yx\left( \Lambda
y^{3}+6\right) }.
\end{equation}
indicating that there are quantum corrections in this case for and the emergent spacetime will be different from the initial Szekeres system with cosmological constant.
\section{Conclusions}
In this work, we investigated the classical and quantum dynamics and stability of the classical equations of the Szekeres system in the presence of a cosmological constant. Our approach was based on looking for Lie-point symmetries of the reduced system of differential equations \eqref{second_order}. An interesting feature is that these equations are related to a variational principle with Lagrange function \eqref{reduced_lag}. Contrary to what usually happens in gravity, the system we studied was not constrained. This is due to the fact that our Lagrangian is related to the reduced system \eqref{second_order} instead of the initial gravitational system. This feature allowed us to define a probability density, which otherwise it is highly non-trivial for generally covariant systems such as the gravitational ones. It was possible to determine new conservation laws, which indicate that the Szekeres system with the cosmological constant term is a Liouville integrable system. Furthermore, for the classical dynamics we performed a study on the stationary points from where we found that the unique attractor is the de Sitter universe. 

We found exact solutions for the wave function of our models by the use of quadratic as well as linear integrals of motion; these latter exist only when the energy density of the classical system vanishes. To extract the physical meaning of these solutions, we employed two methods: i) we studied the extrema of the probability density without assuming any normalization property, since in quantum cosmology, more generally in quantum gravity, probability is not normalizable due to the presence of the constraints; ii) we used the formalism and interpretation of Bohm-de Broglie theory. This analysis indeed revealed the existence of quantum corrections at the semi-classical limit for certain cases. Here, we mean by semi-classical limit that the conjugate momentum is given by \eqref{semiclass}, even though the Hamilton-Jacobi has been modified by the quantum potential. It is at this limit that the Born rule is assumed to hold \cite{Bohm:1951xx,Bohm:1951xw}.

A note is pertinent here for the use of Bohm-de Broglie theory. This theory constitutes a proper alternative to ordinary quantum theory since it does not lie on the notion of an external observer or the measurement process and can describe a single system. Thus it is applicable to closed quantum systems, like the universe for instance. The presence of an additional term, the quantum potential, in the Hamilton-Jacobi equation indicates whether there are quantum effects and thus deviations from the classical solution. This is indeed the case for several cases we studied. Still though there is the question of its compatibility with special and general relativity since this theory is not fundamentally Lorentz covariant due to nonlocality inserted by the quantum potential. This issue is still open and several resolutions have been proposed. One proposal is that Lorentz covariance is emergent and that appears when the quantum potential vanishes \cite{bohm2006undivided}; a second attempt is to covariantly determine the preferred foliation by the wave function \cite{D_rr_2014}; while another is that in the context of quantum cosmology there is no actual problem with the choice of a preferred foliation and thus a global time \cite{10.2307/193027}. Even though in our analysis these foundational issues enter, it is out of the scope of the paper to discuss these open problems.

The analysis for the probability density which follows from the quantum quadratic integrals revealed that the extrema of the probability appear at the stationary points of the classical system. This indicates that there are no quantum corrections in this case. Moreover, this was verified through the calculation of the quantum potential, a term which appears in the gravitational Hamilton-Jacobi equation and originates from quantum effects. Therefore, in this and other cases we apply the linear symmetries, the spacetime becomes purely the well-known classical one.

In conclusion, this analysis contributes to the discussion on the classical and quantum integrability of inhomogeneous systems and extends a previous study in the literature. In a future work we plan to investigate the quantization process in other inhomogeneous cosmological or more complicated gravitational models, which do not exhibit many symmetries.
\bibliographystyle{spphys}       
\bibliography{szekeres}
\end{document}